\documentclass[10pt,aps,twocolumn,notitlepage]{revtex4-2}
\usepackage{amsmath,amssymb,graphicx,bm}
\usepackage[english]{babel}
\usepackage[utf8]{inputenc}
\usepackage{graphicx}
\usepackage{hyperref}
\hypersetup{
    colorlinks,
    linkcolor={red!50!black},
    citecolor={blue!50!black},
    urlcolor={blue!80!black}
}
\usepackage{xcolor}
\usepackage[normalem]{ulem}
\usepackage{float}
\usepackage{comment}
\usepackage{tikz}
\usepackage[a4paper,margin=1in]{geometry}

\newcommand{\Imat}[1]{\mathbf{I}_{#1}}
\newcommand{\Wmat}{\mathbf{W}}
\newcommand{\avec}{\mathbf{a}}
\newcommand{\uvec}{\mathbf{u}}
\newcommand{\Ocal}{\mathcal{O}}

\begin{document}

\title{High-Order Matrix Numerov for Singular Potentials}

\author{Nir Barnea}
\affiliation{The Racah Institute of Physics, The Hebrew University, 91904,
Jerusalem, Israel}

\date{\today}

\begin{abstract}
The matrix Numerov method provides an efficient framework for solving the time-independent Schr\"odinger equation as a matrix eigenvalue problem. However, for singular potentials such as the Coulomb interaction, the expected fourth-order convergence deteriorates for low angular momenta due to the behavior of the potential near the origin. We show that this loss of accuracy originates from an implicit boundary assumption in the standard formulation. By incorporating analytic near-origin information into the discretized Hamiltonian, we derive simple boundary corrections that restore fourth-order convergence and can even produce higher convergence rates for $s$- and $p$-wave energies. The resulting scheme preserves the simplicity and computational efficiency of the original method while significantly improving its accuracy for singular potentials.
\end{abstract}

\maketitle
\section{Introduction}

The time-independent Schr\"odinger equation plays a central role in quantum mechanics, yet analytic solutions are available only for a limited class of model potentials~\cite{LandauQM,GriffithsQM}. For general interactions, numerical methods are therefore indispensable. Among finite-difference approaches, the Numerov method is particularly attractive due to its fourth-order accuracy and its simplicity when applied to second-order differential equations that do not contain first-derivative 
terms~\cite{BLATT1967,Johnson1977}. Because of its efficiency and accuracy, the method has become a standard tool in many quantum-mechanical applications.

The matrix Numerov (MN) method, introduced by Pillai, Goglio, and Walker \cite{PGW12}, 
reformulates the traditional Numerov scheme as a generalized matrix eigenvalue problem. 
This representation allows one to compute bound-state spectra using standard linear 
algebra routines while retaining fourth-order convergence. 
The method is compact, easy to implement, and well suited for both 
pedagogical and 
research applications, see e.g.
\cite{Awasthi2024,Bagci2022,Kaushal25}. 

However, as we noticed while attempting to use the method as a teaching aid
in an undergraduate quantum-mechanics course, when applied to singular potentials -- most notably the Coulomb interaction
\begin{equation}
  V(r) = -\frac{Z}{r},
\end{equation}
the expected fourth-order convergence is not uniformly achieved. In particular, for hydrogenic systems the convergence order is reduced to second order for $s$-waves and to third order for $p$-waves, while the expected fourth-order behavior is recovered only for angular momenta $\ell \ge 2$. This slow convergence was already noted in Ref.~\cite{PGW12}, where the hydrogen atom was identified as the most challenging example.

In this work we show that the reduced convergence originates from an implicit boundary assumption in the standard MN derivation. Specifically, the formulation effectively assumes a regular behavior of the effective potential term at the origin that is incompatible with the $1/r$ singularity of the Coulomb interaction and the $1/r^2$ singularity of the centrifugal term, as well as with the corresponding near-origin structure of the radial wavefunction.

By incorporating analytic information about the wavefunction near $r=0$ directly into the discretized Hamiltonian, we derive simple boundary corrections that restore fourth-order convergence for the $\ell=0$ and $\ell=1$ spectra and may even yield higher-order convergence. The resulting scheme preserves the conceptual and computational simplicity of the original matrix Numerov method while achieving higher accuracy.

For completeness, we briefly review the finite-difference and Numerov formulations before introducing the modified boundary treatment and presenting numerical results.
\section{The Finite Difference Method}

We consider the time-independent radial Schr\"odinger equation for a 
particle with 
angular momentum $\ell$ moving in a central potential $V(r)$. 
In natural units, the equation reads
\begin{equation}\label{Schroedinger}
-\frac{1}{2} u''(r) + W(r)u(r) = \varepsilon u(r),
\end{equation}
where
\begin{equation}
W(r) = V(r) + \frac{\ell(\ell+1)}{2r^2},
\end{equation}
is the effective potential.

We discretize the radial coordinate on a uniform grid with spacing $\Delta$, 
such that $ r_k = k\Delta,$ and $k = 0,1,2,\dots,N $. 
We denote $u_k=u(r_k)$ and similarly $W_k=W(r_k)$.

Using the Taylor expansion
\begin{align}\label{Taylorupm}
  \frac{1}{2}\Big[u(r&+\Delta) + u(r-\Delta)\Big]
  \cr &=
  u(r)
  + \frac{\Delta^2}{2!}u''(r)
  + \frac{\Delta^4}{4!}u^{(4)}(r)
  + \dots
\end{align}
we obtain the standard second-derivative approximation
\begin{equation}\label{upp}
  u''_k
  =
  \frac{u_{k+1} - 2u_k + u_{k-1}}{\Delta^2}
  + \mathcal{O}(\Delta^2).
\end{equation}
Substitution into Eq.~\eqref{Schroedinger} yields the algebraic equation
\begin{equation}
  -\frac{1}{2}\frac{u_{k+1} - 2u_k + u_{k-1}}{\Delta^2}
  +
  W_k u_k
  =
  \varepsilon u_k,
\end{equation}
subject to the boundary conditions $u_0=u_{N+1}=0$.

This can be written as the matrix eigenvalue problem
\begin{equation}
  H \mathbf{u} = \varepsilon \mathbf{u},
\end{equation}
with 
\begin{equation}
   H = -\frac{1}{2\Delta^2}\left(\Imat{+1}-2\Imat{0}+\Imat{-1}\right)
   +\Wmat~.
\end{equation}
Here $\Wmat=\text{diag}(W_1,\ldots W_k, \ldots W_N)$ is the effective potential 
matrix, and $\Imat{p}$ denotes a matrix with ones on the $p$th diagonal, 
and zeros elsewhere. 
As can be seen in \eqref{upp}, the expected accuracy of the method
is $\Ocal({\Delta^2})$.
\section{Numerov Method}
The Numerov method improves the accuracy of the finite difference method
by incorporating higher-order terms.
Utilizing the second derivative of the Taylor expansion \eqref{Taylorupm},
\begin{align}
  \frac{1}{2}\Big[&u''(r+\Delta) + u''(r-\Delta)\Big]=
  \cr &=
  u''(r)
  + \frac{\Delta^2}{2!}u^{(4)}(r)
  + \frac{\Delta^4}{4!}u^{(6)}(r)
  + \dots
\end{align}
we eliminate the fourth-order term $\Delta^4 u^{(4)}$ and obtain
\begin{align}\label{upp_numerov}
  & \frac{u_{k+1} - 2u_k + u_{k-1}}{\Delta^2} =
  \cr & \hspace{1em} = 
  \frac{1}{12}
  \left(  u''_{k+1}  + 10u''_k  + u''_{k-1}  \right)
  +
  \mathcal{O}(\Delta^4).
\end{align}
Now, using the Schr\"odinger equation 
$ u''(r) = 2\left[W(r)-\varepsilon\right]u(r)$,
we can write
\begin{align}\label{numerov}
  &\frac{u_{k+1} - 2u_k + u_{k-1}}{2\Delta^2} =
  \cr & \hspace{2em} =
  \frac{1}{12}
  \left[ W_{k+1}u_{k+1} + 10W_k u_k + W_{k-1}u_{k-1} \right]
  \cr & \hspace{2em} 
  -
  \frac{\varepsilon}{12}
  \left[u_{k+1} + 10u_k + u_{k-1} \right].
\end{align}
This can be written in matrix form 
\begin{equation}\label{mnm}
  H \mathbf{u} = \varepsilon B \mathbf{u},
\end{equation}
or
\begin{equation}\label{mnm}
  B^{-1}H \mathbf{u} = \varepsilon \mathbf{u},
\end{equation}
with
\begin{equation}
   H = -\frac{1}{2\Delta^2}\left(\Imat{+1}-2\Imat{0}+\Imat{-1}\right)
   + B\Wmat,
\end{equation}
and
\begin{equation}
   B = \frac{1}{12}\left(\Imat{+1} + 10\Imat{0} + \Imat{-1}\right).
\end{equation}

Eliminating the leading $\Ocal(\Delta^2)$ error term, 
Eq. \eqref{upp_numerov}, 
the expected accuracy of the Numerov method is $\Ocal({\Delta^4})$.

\section{The singularity at the origin}
\label{sec:singularity}
The derivation of the Numerov method hides a subtle assumption. 
Stepping from Eq. \eqref{upp_numerov} to
\eqref{numerov} it is implicitly assumed that $W_0u_0=0$.
Recalling that asymptotically when $r\to 0$ the wave-function
$u\propto r^{l+1}$, this assumption is clearly invalid for $s$-waves
($\ell=0$)
when the potential contains a Coulomb like $1/r$ singularity, 
and for $p$-waves ($\ell=1$) due to the centrifugal $\ell(\ell+1)/2r^2$ term.

Equiped with this observation, we understand that  
we must modify the Hamiltonian matrix at the origin.
Inspecting Eqs. \eqref{upp_numerov}, and \eqref{numerov} we conclude
that this ammounts to changing the first line in the Hamiltonian matrix
in the following way: 
\begin{equation}
H_{k,k'} \;\longrightarrow\; H_{k,k'} + \frac{1}{24}\delta_{k1}\,a_{k'}.
\end{equation}
Here $\avec$ should be regarded a raw vector containing the numerical 
representation of $u''$ at the origin, i.e. $u''_0=\avec \cdot \uvec$.

In the following subsections we will derive the values of $\avec$
for the different partial waves $\ell$. To this end we assume that
the potential contains a Coulomb like singularty, and that 
it can be expanded near the origin as
\begin{equation}
  V(r) \approx -\frac{Z}{r}+V_0 + V_1 r+\ldots ~.
\end{equation}
\subsection{The $\ell = 0$ case}
\label{ModifiedMNl0}
For $\ell = 0$, 
the wave-function expansion near the origin takes the form
\begin{equation}\label{Taylor0}
  u(r) = u_0' r 
  + \frac{1}{2} u_0'' r^2
  + \frac{1}{3!} u_0^{(3)} r^3
  + \cdots
\end{equation}
Substituting this expansion into the Schr\"odinger equation
\eqref{Schroedinger} 
we obtain at $r=0$
\begin{equation}
  u_0'' = -2Z u_0'.
\end{equation}
Using the Taylor epansion \eqref{Taylor0} we can relate $u_k$ -- 
the wave-function at a grid point $k$ -- to its derivatives
at the origin. Specifically, for the first three grid points we can write
\begin{align}
  u_1 &= u_0' \Delta
  + \frac{1}{2} u_0'' \Delta^2
  + \frac{1}{3!} u_0^{(3)} \Delta^3
  + \cdots, 
  \cr
  u_2 &= 2u_0' \Delta
  + 2u_0'' \Delta^2
  + \frac{8}{3!} u_0^{(3)} \Delta^3
  + \cdots.
  \cr
  u_3 &= 3 u'_0 \Delta + \frac{9}{2}u_0'' \Delta^2 +\ldots~.
\end{align}
Utilizing these relations we can approximate $u'_0$ at various orders.

\paragraph*{First-Order Approximation - }
Using only $u_1$ we can derive a first order approximation for $u_0'$, 
\begin{equation}
u_0' = \frac{u_1}{\Delta} + \Ocal(\Delta)
\end{equation}
implying that
\begin{equation}
u_0'' = -2Z \frac{u_1}{\Delta} + \Ocal(\Delta).
\end{equation}
This result suggest that the first order correction to the Hamiltonian
matrix, should include a modification of the $H_{11}$ matrix element in 
the following way
\begin{equation}
  H_{11}\to H_{11}^{(1)} = H_{11}-\frac{1}{12}\frac{Z}{\Delta}~.
\end{equation}

\paragraph*{Second-Order Approximation - }
A better approximation can be abtained if we use both $u_1$, and $u_2$
eliminating the second-order terms. In this case we obtain
\begin{equation}
  u_0' = \frac{4u_1 - u_2}{2\Delta}
  + \Ocal(\Delta^2).
\end{equation}
Hence,
\begin{align}
  u_0''&=
  - \frac{4Z}{\Delta} u_1
  + \frac{Z}{\Delta} u_2 + \Ocal(\Delta^2).
\end{align}
The resulting second order modification to the Hamiltonian matrix
elements are
\begin{align}
  H_{11}\to H_{11}^{(2)} &= H_{11} - \frac{1}{24}\frac{4Z}{\Delta}, \\
  \\
  H_{12}\to H_{12}^{(2)} &= H_{12} + \frac{1}{24}\frac{Z}{\Delta}.
\end{align}

\paragraph*{Third-Order Approximation - }
We can further improve the approximation of $u_0''$ by considering also
the third point $u_3$.
Extracting $u_0'$ from the three equations for $u_1,u_2,u_3$ we obtain 
the following
modification of the Hamiltonian matrix elements
\begin{align}
  H_{11}^{(3)} &= H_{11} - \frac{1}{12}\frac{3Z}{\Delta},
  \cr
  H_{12}^{(3)} &= H_{12} + \frac{1}{12}\frac{3Z}{2\Delta},
  \cr
  H_{13}^{(3)} &= H_{13} - \frac{1}{12}\frac{Z}{3\Delta}.
\end{align}
\subsection{The $\ell = 1$ Case}
\label{ModifiedMNl1}

For $\ell = 1$, the centrifugal term $\ell(\ell+1)/(2r^2)$ 
dominates the singular behavior of the effective potential 
at the origin $r \to 0$. 
Consequently, the radial wavefunction behaves near the origin as
\begin{equation}
  u(r) =
  \frac{1}{2} u_0'' r^2
  + \frac{1}{3!} u_0^{(3)} r^3
  + \cdots .
\end{equation}

Evaluating this expansion at the first grid point, $r_1=\Delta$, yields
\begin{equation}
  u_1 =
  \frac{1}{2} u_0'' \Delta^2
  + \frac{1}{3!} u_0^{(3)} \Delta^3
  + \cdots .
\end{equation}
Retaining only the leading contribution gives
\begin{equation}
  u_0'' = \frac{2u_1}{\Delta^2} + \Ocal(\Delta).
\end{equation}

Substituting this result into the boundary correction term, 
we obtain the modified Hamiltonian matrix element
\begin{equation}
  H_{11} \to H_{11}^{(1)}
  =
  H_{11} + \frac{1}{12}\frac{1}{\Delta^2}.
\end{equation}

\paragraph*{Second-Order Approximation - }
As for the $\ell=0$ case we can further improve the accuracy considering
also the second point $u_2$. 
Estimating $u_0''$ at this order we get
\begin{equation}
  u_0'' = \frac{4u_1}{\Delta^2} - \frac{u_1}{2\Delta^2} +\Ocal(\Delta^2).
\end{equation}
The resulting modification of the Hamiltonian matrix elements are
\begin{align}
  H_{11}^{(2)} &= H_{11} + \frac{1}{12}\frac{2}{\Delta^2},
  \cr
  H_{12}^{(2)} &= H_{12} - \frac{1}{12}\frac{1}{4\Delta^2}.
\end{align}
\subsection{The $\ell \ge 2$ Case }
For $\ell \ge 2 $ there are no corrections to the Hamiltonian matrix.
In these case the $u\propto r^{l+1}$ behavior of the wave-function at the 
origin suffices to cancel the singularity in the effective potential.

\section{Numerical Results}

To validate the modified matrix Numerov scheme, we consider two benchmark problems: a particle moving in a harmonic oscillator potential and the hydrogen atom. For each system we study the convergence of the energy levels $E_{n\ell}$ associated with the radial quantum number $n$ and angular momentum $\ell$, as well as the corresponding radii $R_{n\ell}$ defined through
\begin{equation}
R_{n\ell}^2 = \int_0^\infty dr\, u_{n\ell}^2(r)\, r^2 .
\end{equation}

To quantify the convergence of the numerical solutions we introduce the absolute deviations
\begin{equation}
\Delta E_{n\ell}(N) = |E_{n\ell}(N) - E^{\text{Theo}}_{n\ell}| ,
\end{equation}
\begin{equation}
\Delta R_{n\ell}(N) = |R_{n\ell}(N) - R^{\text{Theo}}_{n\ell}| ,
\end{equation}
where $N$ is the number of grid points and $E^{\text{Theo}}_{n\ell}$ and $R^{\text{Theo}}_{n\ell}$ denote the exact values. 

The numerical results are fitted to the scaling form
\begin{equation}\label{fitq}
\Delta E_{n\ell}(N) = C N^{-q},
\end{equation}
and similarly for the radius. The exponent $q$ therefore measures the effective convergence order of the method. For the standard finite-difference discretization we expect $q \approx 2$, whereas the Numerov method should exhibit $q \approx 4$ provided the fourth-order accuracy is preserved.

\subsection{Harmonic Oscillator}

As a first test we consider a particle moving in a three-dimensional harmonic oscillator potential
\begin{equation}
V(r) = \frac{1}{2} r^2 .
\end{equation}
Since this potential is regular at the origin, it provides a useful baseline for assessing whether the reduced convergence observed for singular potentials originates from the potential itself or from the discretization procedure.

The exact energy levels and corresponding radii are given (in natural units) by
\begin{equation}
E_{n\ell} = 2n + \ell + \frac{3}{2}, 
\quad 
R_{n\ell} = \sqrt{2n + \ell + \frac{3}{2}},
\end{equation}
for integer $n,\ell \ge 0$.

Table~\ref{ho_tabq} summarizes the extracted convergence exponents $q$ (rounded to the nearest integer) for the energies and radii of the states $n=0$, $\ell=0,1,2$. As expected, the finite-difference method exhibits second-order convergence for all calculated quantities. The standard matrix Numerov method reproduces the expected fourth-order convergence for most cases, with the exception of the $\ell=1$ energy where the observed convergence rate is $q=3$.

\begin{figure}\begin{center}
\includegraphics[width=\linewidth]{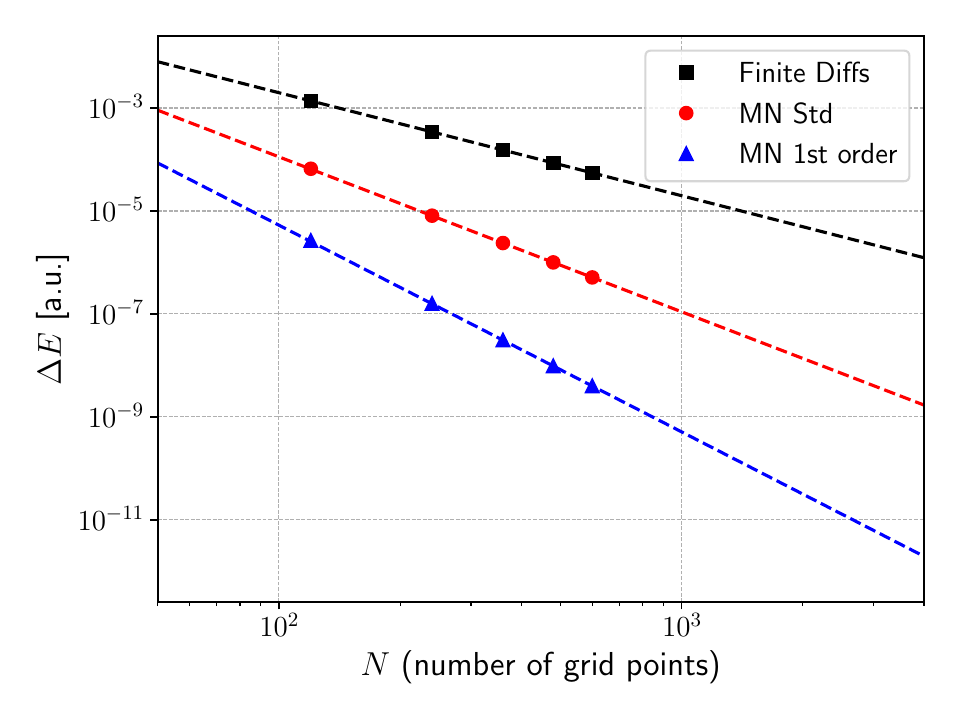}
\caption{\label{fig:HO_ENR_L1} 
Convergence rate of the harmonic oscillator energy level $n=0,\ell=1$ 
as a function of the number of grid points.
Red circles - the standard matrix Numerov method \cite{PGW12},
blue triangles - Matrix Numerov with first order correction,
black squares - the finite difference method.
Dashed lines - fit to equation \eqref{fitq}.
}
\end{center}
\end{figure}

Incorporating the first-order boundary correction introduced for the $\ell=1$ case in Sec.~\ref{ModifiedMNl1} substantially improves the energy convergence, restoring fourth-order behavior. This effect is illustrated in
Fig.~\ref{fig:HO_ENR_L1}.

\begin{table}
\begin{center}
\caption{\label{ho_tabq} Harmonic Oscillator - the convergence exponent 
$q$ (rounded) of the energies $E$ and radii $R$ of the $n=0$, $\ell=0,1,2$ states.}
\begin{tabular}{|c| c | c | c | c | c | c |}
\hline\hline
 & \multicolumn{2}{c|}{$\ell=0$}
 & \multicolumn{2}{c|}{$\ell=1$}
 & \multicolumn{2}{c|}{$\ell=2$} \\
\hline
 &  $\hspace{0.5em} E \hspace{0.5em}$ & $\hspace{0.5em} R \hspace{0.5em}$ 
 &  $\hspace{0.5em} E \hspace{0.5em}$ & $\hspace{0.5em} R \hspace{0.5em}$ 
 &  $\hspace{0.5em} E \hspace{0.5em}$ & $\hspace{0.5em} R \hspace{0.5em}$  \\
\hline
Finite Diffs   
 & $2$ & $2$ & $2$ & $2$ & $2$ & $2$ \\
\hline
MN - Std 
 & $4$ & $4$ & $3$ & $4$ & $4$ & $4$ \\
\hline
MN 1$^{st}$-order 
 & $4$ & $4$ & $4$ & $4$ & $-$ & $-$ \\
\hline
\end{tabular}
\end{center}
\end{table}

\subsection{The Hydrogen Atom}

We next consider the hydrogen atom, where the electron moves in the attractive Coulomb potential
\begin{equation}
V(r) = -\frac{1}{r}.
\end{equation}

The exact energy spectrum is given by
\begin{equation}
E_{n\ell} = -\frac{1}{2(n+\ell)^2},
\end{equation}
with $n \ge 1$. The corresponding radii are
\begin{equation}
R_{n\ell} =
\sqrt{
\frac{(n+\ell)^2}{2}
\left[5(n+\ell)^2 + 1 - 3\ell(\ell+1)\right]
}.
\end{equation}

Table~\ref{hatom_tabq} presents the extracted convergence exponents $q$ for the hydrogen atom. The standard matrix Numerov method reproduces the expected fourth-order convergence only for angular momenta $\ell \ge 2$. For $\ell=1$ the convergence is reduced to third order, while for $\ell=0$ it drops to second order. This behavior is fully consistent with the analysis presented in Sec.~\ref{sec:singularity}.

Incorporating the first-order boundary correction significantly improves the convergence for both energy and radius, raising the convergence order to $q=3$ for $\ell=0$ and restoring fourth-order convergence for $\ell=1$. Higher-order boundary corrections further improve the $s$-wave convergence. 
For $\ell=0$ the second-order correction yields $q=4$, while surprisingly 
the third-order correction increases the convergence rate to $q\approx 5$.

\begin{figure}\begin{center}
\includegraphics[width=\linewidth]{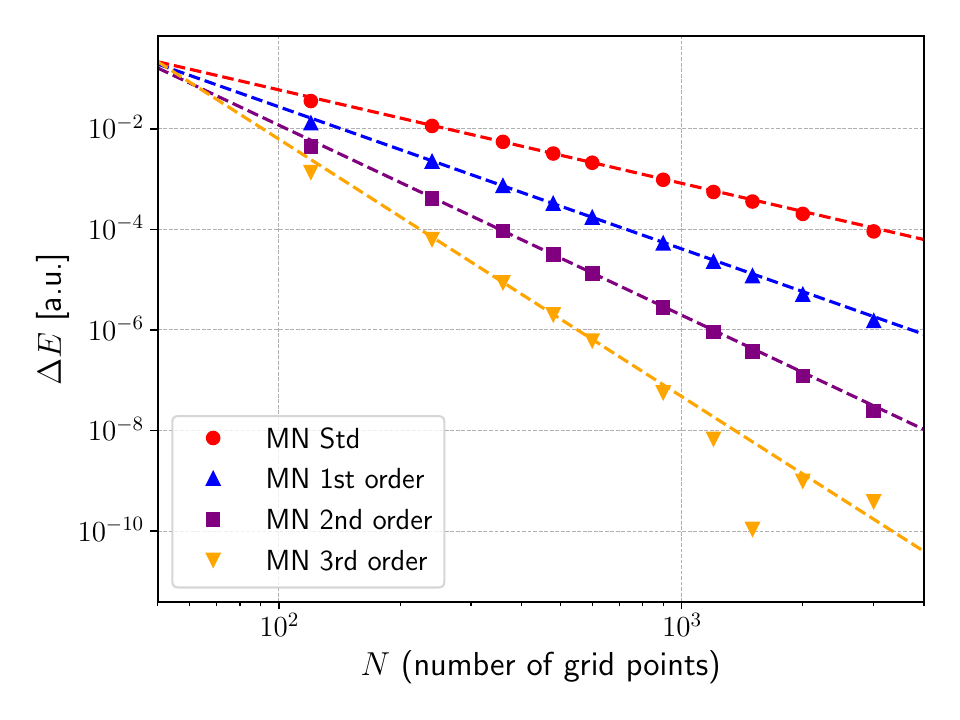}
\caption{\label{fig:L0} 
Convergence rate of the hydrogen ground state energy
$n=1,\ell=0$ as a function of the number of grid points.
Red circles - the original matrix Numerov method \cite{PGW12},
blue triangles - Matrix Numerov with first order correction,
purple squares - with second order corrections,
orange triangles - with third order corrections.
Dashed lines - fit to equation \eqref{fitq}.
}
\end{center}
\end{figure}

\begin{figure}\begin{center}
\includegraphics[width=\linewidth]{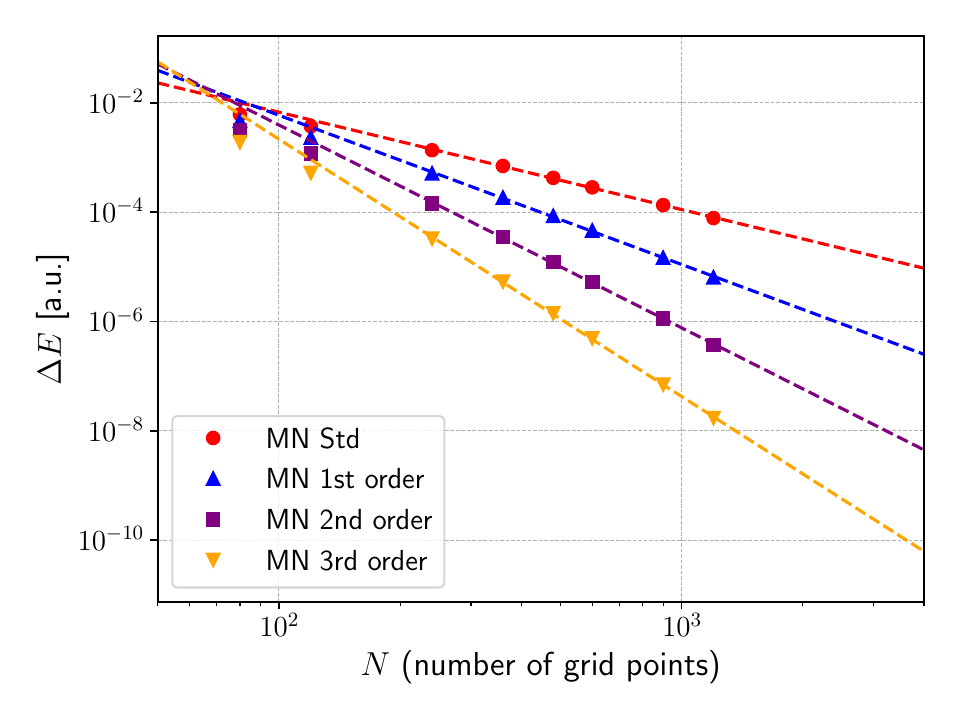}
\caption{\label{fig:N2L0} 
Convergence rate of the hydrogen energy level
$n=3,\ell=0$ as a function of the number of grid points.
Red circles - the original matrix Numerov method \cite{PGW12},
blue triangles - Matrix Numerov with first order correction,
purple squares - with second order corrections,
orange triangles - with third order corrections.
Dashed lines - fit to equation \eqref{fitq}.
}
\end{center}
\end{figure}

This convergence pattern is illustrated in Fig.~\ref{fig:L0} for the hydrogen ground state $(n=1,\ell=0)$ and in Fig.~\ref{fig:N2L0} for the excited state $(n=3,\ell=0)$. The slope of the log--log plots clearly demonstrates the transition from second-order convergence in the standard matrix Numerov method to fifth-order scaling when the third-order analytic boundary treatment is included. Figure~\ref{fig:N2L0} further shows that the convergence rates summarized in Table~\ref{hatom_tabq} also hold for excited states.

\begin{figure}\begin{center}
\includegraphics[width=\linewidth]{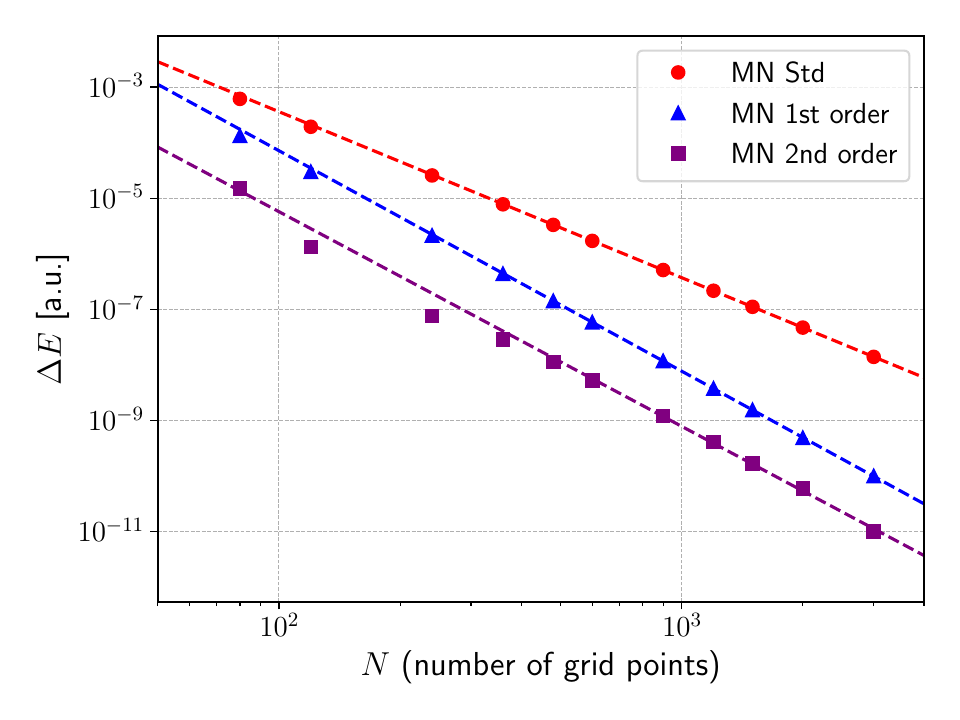}
\caption{\label{fig:L1} 
Convergence rate of the hydrogen energy level
$n=1,\ell=1$ as a function of the number of grid points.
Red circles - the original matrix Numerov method \cite{PGW12},
blue triangles - Matrix Numerov with first order correction,
purple squares - with second order corrections.
Dashed lines - fit to equation \eqref{fitq}.
}
\end{center}
\end{figure}

For $\ell=1$, incorporating second-order corrections does not change the asymptotic convergence exponent $q$, but it nevertheless improves the overall numerical accuracy. This behavior is illustrated in Fig.~\ref{fig:L1}.

Taken together, these numerical results demonstrate that the reduced convergence of the standard matrix Numerov method for Coulomb potentials originates almost entirely from the discretization near the origin. Incorporating analytic boundary information dramatically improves the accuracy of the method while preserving its simplicity and computational efficiency.

\begin{table}
\begin{center}
\caption{\label{hatom_tabq} The hydrogen atom - the convergence exponent 
$q$ (rounded) of the energies $E$ and radii $R$ of the $n=1$, $\ell=0,1,2$ states.}
\begin{tabular}{|c| c | c | c | c | c | c |}
\hline\hline
 & \multicolumn{2}{c|}{$\ell=0$}
 & \multicolumn{2}{c|}{$\ell=1$}
 & \multicolumn{2}{c|}{$\ell=2$} \\
\hline
 &  $\hspace{0.5em} E \hspace{0.5em}$ & $\hspace{0.5em} R \hspace{0.5em}$ 
 &  $\hspace{0.5em} E \hspace{0.5em}$ & $\hspace{0.5em} R \hspace{0.5em}$ 
 &  $\hspace{0.5em} E \hspace{0.5em}$ & $\hspace{0.5em} R \hspace{0.5em}$  \\
\hline
Finite Diffs   
 & $2$ & $2$ & $2$ & $2$ & $2$ & $2$ \\
\hline
MN - Std 
 & $2$ & $2$ & $3$ & $3$ & $4$ & $4$ \\
\hline
MN 1$^{st}$-order 
 & $3$ & $3$ & $4$ & $4$ & $-$ & $-$ \\
\hline
MN 2$^{nd}$-order 
 & $4$ & $4$ & $4$ & $4$ & $-$ & $-$ \\
\hline
MN 3$^{rd}$-order 
 & $5$ & $5$ & $-$ & $-$ & $-$ & $-$ \\
\hline
\hline
\end{tabular}
\end{center}
\end{table}

\section{Conclusions}

We have analyzed the loss of fourth-order convergence of the matrix Numerov method in the presence of Coulomb and centrifugal singularities and identified its origin in an implicit boundary assumption at the origin. While for regular potentials and for angular momenta $\ell \ge 2$ the standard formulation achieves the expected $O(\Delta^4)$ accuracy, the convergence deteriorates to second order for $s$-waves when a $1/r$ interaction is present, and to third order for $p$-waves regardless the form of the potential.

By incorporating analytic near-origin information for the radial 
wavefunction directly into the discretized Hamiltonian, 
we derived simple boundary corrections that improve the asymptotic 
convergence rate for $\ell=0$ and $\ell=1$. 
The modification affects only the first row of the Hamiltonian matrix 
and therefore preserves the structure and computational efficiency of the 
original matrix Numerov scheme.

The resulting method provides a transparent and systematic way to 
treat Coulomb-type and centrifugal singularities within the matrix 
Numerov framework. 
It is therefore well suited for high-precision calculations of 
hydrogenic systems and can be straightforwardly extended to other 
central potentials with similar singular behavior.

\section*{Generative AI}
Statement: During the preparation of this work the author used ChatGPT 5.2
as a text and language editor. After using this tool/service, the author reviewed and edited the content as needed and takes full responsibility for the content of the published article.

\begin{acknowledgments}
This research was supported by 
the Israel Science Foundation under grant number ISF 2441/24.
\end{acknowledgments}
\bibliographystyle{apsrev4-2}
\bibliography{NMBib}  
\end{document}